\documentclass{PoS}

\def\eq#1{Eq.~(\ref{#1})}
\newcommand{\secn}[1]{Section~\ref{#1}}

\newcommand{\ord}{{\cal O}}
\newcommand{\RE}{{\rm Re}}

\def\beq{\begin{equation}}
\def\eeq{\end{equation}}
\def\beqa{\begin{eqnarray}}
\def\eeqa{\end{eqnarray}}
\def\ifm{\ifmmode}

\def \as {\relax\ifmmode\alpha_s\else{$\alpha_s${ }}\fi}

\def \al #1 {\frac {\as({#1})}{\pi} }
\def \ds #1 {\ooalign{$\hfil/\hfil$\crcr$#1$}}

\def\eps{\varepsilon}
\def\eq#1{Eq.~(\ref{#1})}

\title{Beyond Reggeization for two- and three-loop QCD amplitudes}

\ShortTitle{Beyond Reggeization}

\author{Vittorio Del Duca\\
            INFN, Laboratori Nazionali di Frascati\\
            E-mail: \email{delduca@lnf.infn.it}}

\author{Giulio Falcioni\\
        University of Torino and INFN, Sezione di Torino\\
        E-mail: \email{falcioni@to.infn.it}}

\author{\speaker{Lorenzo Magnea}\\
        University of Torino and INFN, Sezione di Torino\\
        E-mail: \email{lorenzo.magnea@unito.it}}

\author{Leonardo Vernazza\\
        INFN, Sezione di Torino\\
        E-mail: \email{vernazza@to.infn.it}}

\abstract{The high-energy factorization of gauge theory scattering amplitudes in terms of 
universal impact factors and a Reggeized exchange in the $t$-channel, corresponding to 
a Regge pole in the angular momentum plane, is know to conflict with the structure of 
soft anomalous dimensions starting at the two-loop level. We explore the implications
of this violation of factorization for two- and three-loop QCD amplitudes: first we propose
criteria to organize the amplitudes into factorizing and non-factorizing terms, then we
test them by recovering a known result for non-logarithmic terms at two loops. Finally
we predict the precise value of the leading non-factorizing energy logarithms at three 
loops, and we uncover a set of all-order identities constraining infrared finite terms in
quark and gluon amplitudes.}

\FullConference{11th International Symposium on Radiative Corrections (Applications of 
                           Quantum Field Theory to Phenomenology) (RADCOR 2013),\\
		          22-27 September 2013\\
		          Lumley Castle Hotel, Durham, UK }

\begin{document}

\section{Introduction}
\label{intro}

Recent years have seen truly remarkable progress in our understanding 
of scattering amplitudes in quantum field theory. New techniques have 
been developed  to compute scattering amplitudes of increasing complexity 
at finite orders in perturbation theory (see for example the recent reviews
\cite{Elvang:2013cua,Dixon:2013uaa}), and new approaches, emerging
from special theories, are introducing radically different points of 
view~\cite{ArkaniHamed:2012nw}. Within this fast-developing framework,
general properties of amplitudes in special kinematical regimes, notably
factorization and universality, continue to provide insights which apply to
all orders in perturbations theory, and which prove useful both for testing
high-order calculations, and for direct phenomenological applications.

Our goal in this contribution is to examine gauge theory scattering amplitudes
using constraints that arise from two different kinematical limits, where two
different factorizations, and two different notions of universality, are known 
to apply. Our main target is the calculation of amplitudes in the high-energy
limit, characterized by the fact that the center-of-mass energy $\sqrt{s}$
is much larger than all other kinematic invariants. In this limit, large logarithms 
of $s$ dominate the perturbative expansion and, at least to some finite logarithmic 
accuracy, they can be resummed using Regge theory~\cite{Collins:1977jy}.
While the validity of Regge theory rests upon very general principles of 
quantum field theory, and indeed many related results predate and anticipate
subsequent developments in QCD, concrete applications to the perturbative
resummation of energy logarithms have been so far confined to leading and
next-to-leading logarithms. Our main tool to go beyond existing results for
the high-energy limit will be soft-collinear factorization: the statement that
leading contributions to scattering amplitudes from virtual particles carrying
vanishing energy, or with velocities collinear to the external momenta, can
be factorized and thus resummed in exponential form. Soft-collinear factorization
is valid to all orders in perturbation theory, but of course applies only to 
divergent parts of the amplitude.

It is not surprising that information on the high-energy limit can be gained by
considering the infrared properties of scattering amplitudes: in the presence
of some kinematic cutoff, infrared divergences appear as logarithmic enhancements
involving the ratio of the characteristic energy scale of the process with the 
new scale provided by the cutoff: in this sense, high-energy logarithms can 
be seen to some extent as a special class of infrared logarithms. This 
kind of insight has been used to explore the high-energy limit since early 
days~\cite{Korchemsky:1993hr,Korchemskaya:1994qp,Korchemskaya:1996je}. 
More recently, a detailed analysis of soft-collinear 
factorization in the high-energy limit~\cite{DelDuca:2011ae,Bret:2011xm}
showed that the simplest form of Regge factorization, which is based on the 
approximation of considering only Regge poles as possible singularities in the
angular momentum plane, breaks down for infrared divergent contributions at 
the level of next-to-next-to leading logarithms (NNLL). Conversely, and even 
more recently, Ref.~\cite{Caron-Huot:2013fea} used constraints derived from
Regge factorization to provide evidence for new structures arising at four 
loops in soft anomalous dimensions.

In this note, we will start from the results of~\cite{DelDuca:2011ae,Bret:2011xm}
and apply them to the explicit calculation of finite order quark and gluon amplitudes 
in QCD. We begin, in \secn{tale}, by reviewing the basic features of the two factorizations 
that we shall be employing, and the precise form they take for the case at hand.
Then we will show, in \secn{pertexp}, how finite order constraints can be derived 
by comparing the two factorization, starting with an illustration at the one-loop level.
Our main results~\cite{DelDuca:2013ara} are given in \secn{resu}: there, starting at 
the two-loop level, we show how one can use soft-collinear factorization to detect 
universality-breaking terms at high energy. We organize these terms into remainder 
functions whose divergent contributions can be determined using infrared information, 
and we test our ideas at two loops, for non-logarithmic terms, reproducing the results of
Ref.~\cite{DelDuca:2001gu}, where for the first time a violation of Regge factorization
was detected. We then proceed to study non-Reggeizing energy logarithms at three 
loops, and provide explicit predictions for the divergent parts of their coefficients 
in the case of quark and gluon four-point amplitudes. Such contributions, arising
at three loops in non-planar diagrams, are expected to correspond to Regge cuts
in the angular momentum plane, and could provide a boundary condition for a 
more general high-energy resummation, going beyond conventional Reggeization.
We conclude by showing how high-energy factorization can, in turn, provide 
interesting contraints on the finite parts of soft-collinear factorized amplitudes. 
Our results will be extended to subleading infrared poles and discussed in 
greater detail in Ref.~\cite{us}.

\section{A tale of two factorizations}
\label{tale}

We begin by describing the framework of high-energy factorization. As in the rest of
this note, we will consider the case of four-parton amplitudes in QCD, although the 
formalism is much more general and in particular can treat multi-parton amplitudes 
as well. For four partons, in the limit $s \gg |t|$, the amplitude is dominated by color
octet exchange in the $t$ channel. It takes on a factorized form, and we will take as 
master formula for this factorization the expression
\beqa
  {\cal M}_{ab}^{[8]} \left(\frac{s}{\mu^2}, \frac{t}{\mu^2}, \as \right)
  & = & 2 \pi \alpha_s \, H^{(0),[8]}_{ab}  \,
  \Bigg\{
  C_a \left(\frac{t}{\mu^2}, \as \right)
  \bigg[ A_+ \left(\frac{s}{t}, \as \right) + \, \kappa_{ab} \,
  A_- \left(\frac{s}{t}, \as \right) \bigg]
  C_b \left(\frac{t}{\mu^2}, \as \right) \nonumber \\
  & & \, + \, \, {\cal R}_{ab}^{[8]} \left(\frac{s}{\mu^2}, \frac{t}{\mu^2}, \as \right)
 + \ord \left( \frac{t}{s} \right) \Bigg\} \, .
\label{ReggeFact}
\eeqa
Here the indices $a,b = q,g$ denote the parton species (quark or gluon), and we are 
not displaying (as in the rest of the paper) the dependence on the IR regulator $\epsilon 
= 2 - d/2$. The impact factors $C_{a,b}$ are universal functions, depending only on
the identity of the particles that scatter by exchanging the $t$-channel color octet. The 
finite tree-level amplitude $H_{ab}^{(0),[8]}$ contains the $t$-channel propagator 
pole, while the high-energy logarithms are generated by the Regge factors $A_\pm$, 
which are given by
\beq
  A_\pm \left(\frac{s}{t}, \as \right) \, = \, \left( \frac{- s}{- t} \right)^{\alpha(t)}
  \pm \left( \frac{s}{-t} \right)^{\alpha(t)} \, , 
\label{ReggeStructure}
\eeq
where $\alpha(t)$ is the Regge trajectory for octet exchange, which admits a 
perturbative expansion in powers of $\alpha_s$, with IR divergent coefficients.
The Regge factors have already been properly (anti)symmetrized with 
respect to the exchange $s \leftrightarrow u \sim - s$, which also requires the 
inclusion of the factors $\kappa_{ab}$, compensating for the different symmetry 
properties of quark and gluon color factors; specifically, $\kappa_{gg} = \kappa_{qg} 
= 0$, while $\kappa_{qq} = (4 - N_c^2)/N_c^2$. Finally, in \eq{ReggeFact} we have 
allowed for a remainder function ${\cal R}_{ab}^{[8]}$, designed to collect all 
non-factorizing contributions to the amplitude, at leading power in $|t|/s$. 
High-energy factorization is known to be exact (with vanishing remainder) at 
leading~\cite{Balitsky:1979ap} and next-to-leading~\cite{Fadin:1993wh,Fadin:2006bj} 
logarithmic accuracy (for the real part of the amplitude), while evidence of a 
non-vanishing, non-logarithmic remainder at two loops was uncovered in
Ref.~\cite{DelDuca:2001gu}.

Turning now to soft-collinear factorization, we may start with a general expression
generating infrared divergences of arbitrary multi-parton amplitudes~\cite{Dixon:2008gr,
Becher:2009cu,Gardi:2009qi,Becher:2009qa,Gardi:2009zv}. For $n$ partons
with momenta $p_i$ we may write
\beq 
  {\cal M} \left(\frac{p_i}{\mu}, \as \right) \, = \, {\cal Z} \left(\frac{p_i}{\mu}, \as \right)
  {\cal H} \left(\frac{p_i}{\mu}, \as \right) \, ,
\label{IRfact}
\eeq
where the matrix element ${\cal M}$ and the finite hard part ${\cal H}$ are vectors
in the space of available color tensors, while ${\cal Z}$ is an operator in the same
space, which generates all infrared poles in dimensional regularization. Multiplicative
renormalizability implies that the ${\cal Z}$ operator can be written in terms of a
soft anomalous dimension matrix $\Gamma$ as
\beq
  {\cal Z} \left(\frac{p_i}{\mu}, \as \right) \, = \,  
  {\cal P} \exp \left[ \frac{1}{2} \int_0^{\mu^2} \frac{d \lambda^2}{\lambda^2} \, \,
  \Gamma \left(\frac{p_i}{\lambda}, \alpha_s(\lambda^2) \right) \right] \, .
\label{RGsol}
\eeq
The state of the art concerning the soft anomalous dimension matrix $\Gamma$, for
massless partons, is expressed by the dipole formula~\cite{Becher:2009cu,Gardi:2009qi,
Becher:2009qa,Gardi:2009zv}
\beq
  \Gamma_{\rm dip}  \left(\frac{p_i}{\lambda}, \alpha_s(\lambda^2) \right) \, = \,
  \frac{1}{4} \, \widehat{\gamma}_K \left(\alpha_s (\lambda^2) \right) \,
  \sum_{(i,j)} \ln \left(\frac{- s_{i j}}{\lambda^2} 
  \right) {\bf T}_i \cdot {\bf T}_j \, - \, \sum_i
  \gamma_{J_i} \left(\alpha_s (\lambda^2) \right) \, ,
\label{sumodipoles}
\eeq
where ${\bf T}_i$ is a color insertion operator appropriate to parton $i$, $\widehat{\gamma}_K$
is the cusp anomalous dimension~\cite{Korchemsky:1985xj,Korchemsky:1987wg} with 
the Casimir of the relevant color representation scaled out, and $\gamma_{J_i}$ are 
collinear anomalous dimensions for each external parton. The dipole formula is exact 
at two loops~\cite{Aybat:2006mz}, and at three loops can only receive tightly constrained 
corrections~\cite{Becher:2009qa,Dixon:2009ur,Ahrens:2012qz} whose calculation is
under way~\cite{Gardi:2013saa}. Evidence for such corrections at the four-loop level
was uncovered in~\cite{Caron-Huot:2013fea}.

We now follow Refs.~\cite{DelDuca:2011ae,Bret:2011xm}, and take the high-energy limit
of \eq{IRfact}, using \eq{sumodipoles}. For four-point amplitudes, to leading power in $|t|/s$, 
but to all logarithmic accuracies, the infrared operator ${\cal Z}$ factorizes as
\beq
  {\cal Z} \left(\frac{p_i}{\mu}, \as \right) \, = \, 
  {\cal Z}_{\bf 1,  R} \left(\frac{t}{\mu^2}, \as \right) \, 
  \exp \left( - {\rm i} \, \frac{\pi}{2} \, K \left( \as \right)
  {\cal C}_{\rm tot} \right)
  \widetilde{\cal Z} \left(\frac{s}{t}, \as \right) + \ord \left(\frac{t}{s} \right) \, .
\label{Zfact}
\eeq
The crucial ingredient of \eq{Zfact}, which retains a matrix structure in color space, and
generates all energy logarithms, is the factor
\beq
  \widetilde{{\cal Z}} \left(\frac{s}{t}, \as \right)
  \, = \, \exp \left\{ K( \as )
  \left[ \log \left( \frac{s}{-t } \right) {\bf T}_t^2 + {\rm i} \pi {\bf T}_s^2\right] \right\} \, ,
\label{widetildeZ}
\eeq
where we have introduced the `Mandelstam' combinations of color operators ${\bf T}_t = 
{\bf T}_1 + {\bf T}_3$ and ${\bf T}_s = {\bf T}_1 + {\bf T}_2$ for  the scattering process
$1 + 2 \to 3 + 4$, and we defined
\beq
  K \left( \as \right) = 
  - \frac{1}{4} \int_0^{\mu^2} \frac{d \lambda^2}{\lambda^2} \,
  \hat{\gamma}_K \left( \alpha_s (\lambda^2) \right) \, .
\label{cusp}
\eeq
Furthermore, in \eq{Zfact} we have isolated a divergent phase, defining the Casimir
eigenvalue ${\cal C}_{\rm tot}$ as the sum of the four Casimirs associated to the four 
external partons. The remaining factor, ${\cal Z}_{\bf 1, R}$ is then a real, color 
singlet, energy-independent exponential of the form
\beqa
  {\cal Z}_{\bf 1,  R} \left(\frac{t}{\mu^2}, \as \right) \! & = & \!
  \exp \Bigg\{  \frac{1}{2} \Bigg[ K \left( \as \right) \log \left( \frac{-t}{\mu^2} 
  \right) + D \left( \as \right) \Bigg] {\cal C}_{\rm tot}  
  + \, \sum_{i = 1}^4 B_i \left( \as \right) \Bigg\} \, , 
\label{Z1}
\eeqa
where the functions $D(\as)$ and $B_i(\as)$ are scale integrals similar to \eq{cusp}, and
given explicitly in Ref.~\cite{DelDuca:2011ae}. The most relevant feature of \eq{Z1} for our 
present purposes is that it can be written as a product of factors unambiguously associated 
with each external parton, as
\beq
  {\cal Z}_{\bf 1, R} \left( \frac{t}{\mu^2}, \as \right) 
  \, = \, \prod_{i = 1}^4 {\cal Z}_{\bf 1,  R}^{(i)} \left( \frac{t}{\mu^2}, \as \right) \, .
\label{jetfactors}
\eeq
Such an expression strongly suggests that the factors ${\cal Z}_{\bf 1, R}^{(i)}$ should be
related to the (divergent contributions to) the impact factors $C_a$. As we will see in the 
next sections, this is indeed the case.

\section{Perturbative expansions}
\label{pertexp}

Our task is now in principle straightforward: we must expand the all-order factorized
expressions given in the previous section in powers of the coupling and of the energy
logarithm; comparing the resulting perturbative coefficients will make explicit the 
constraints that each factorization implies for the other one, and will give explicit 
results for the violations of high-energy factorization at NNLL that were described
in Refs.~\cite{DelDuca:2011ae,Bret:2011xm}. For example, we expand each color 
component ${\cal M}^{[j]}$  of the full matrix elements as
\beq
  {\cal M}^{[j]} \left(\frac{s}{\mu^2}, \frac{t}{\mu^2}, \as \right) \, = \, 4 \pi \as \, 
  \sum_{n = 0}^\infty \sum_{i = 0}^n
  \left( \frac{\as}{\pi} \right)^n \ln^i \left( \frac{s}{- t} \right)
  M^{(n), i, [j]} \left( \frac{t}{\mu^2} \right) \, ,
\label{AmpExpansion}
\eeq
and similarly for the remainder functions ${\cal R}_{ab}^{[8]}$, where however
we know that leading contributions must arise at ${\cal O} (\as^2)$, and be 
next-to-next-to leading in the energy logarithm. All functions that do not depend 
on the energy such as impact factors and the singlet operator ${\cal Z}_{\bf 1, R}$ 
are simply expanded in powers of $\as/\pi$.

Let us illustrate the procedure at the one loop level. Soft-collinear factorization
at this level gives the expressions
\beqa
  M^{(1),0} & = & \left[ Z_{1, {\bf R}}^{(1)} + i \pi K^{(1)} \left({\bf T}_s^2  - 
  \frac{1}{2} {\cal C}_{\rm tot} \right) \right] H^{(0)} + H^{(1),0}, \nonumber \\
  M^{(1),1} & = & K^{(1)} {\bf T}_{t}^2 H^{(0)} + H^{(1),1} \, ,
\label{AmpCoeff1}
\eeqa
which are still vector equations in color space, whereas, expanding \eq{ReggeFact}, 
we readily find that the color octet components of the one-loop matrix elements
can be expressed as
\beqa
  M^{(1),0,[8]}_{ab} & = & \left[ C_a^{(1)} + C_b^{(1)}
  - i \frac{\pi}{2} (1 + \kappa_{ab}) \alpha^{(1)} \right] H^{(0),[8]}_{a b} \, , \nonumber \\
  M^{(1),1, [8]}_{ab} & = & \alpha^{(1)} H^{(0), [8]}_{ab} \, .
\label{ReggeCoeff1}
\eeqa
Comparing the two results for the octet component at LL level, we readily verify that the 
one-loop Regge trajectory must be given by
\beq
  \alpha^{(1)} \, = \, \frac{K^{(1)} \left( {\bf T}_t^2 H^{(0)} \right)^{[8]} }{H^{(0),[8]}} 
  + \frac{H^{(1),1,[8]}}{H^{(0),[8]}} \, = \, C_A K^{(1)} + \ord(\eps) \, ,
\label{alpha1}
\eeq
as predicted in~\cite{DelDuca:2011ae,Bret:2011xm}. Note that, to get the second 
equality, we made use of the fact that the tree-level matrix element is pure 
color octet at leading power in $|t|/s$, and we used the fact that $H^{(1),1,[8]} 
= {\cal O} (\epsilon)$. On the other hand, the reality of the Regge trajectory requires
that ${\rm Im} (H^{(1),1,[8]}) = 0$, which is easily verified. More interesting constraints
arise when looking at the NLL terms in Eqns.~(\ref{AmpCoeff1}) and (\ref{ReggeCoeff1}).
Concentrating on imaginary parts, one finds that
\beq
 {\rm Im} \left[ H^{(1), 0, [8]} \right]  \, = \, - \, \frac{\pi}{2} \, (1 + \kappa) \, 
 {\rm Re} \left[ H^{(1), 1, [8]} \right]
 \, + \, \frac{\pi}{2} \, K^{(1)} \left( \Big[ {\cal C}_{\rm tot} - 2 {\bf T}_s^2
 - (1 + \kappa) {\bf T}_t^2 \Big] H^{(0)} \right)^{[8]} \, ,
\label{ImNLL1}
\eeq
Now, soft-collinear factorization requires $H^{(1), 0, [8]}$ to be finite, while $K^{(1)}$
is a pure pole. Consistency requires the vanishing of the matrix element
\beq
  \bigg[ {\cal C}_{\rm tot} - 2 {\bf T}_s^2
 - (1 + \kappa) {\bf T}_t^2 \bigg]_{[8],[8]} \, = \, 0 \, ,
\label{magid}
\eeq
for any combination of representations that can participate in the scattering 
process. Using, for example, the explicit results for the color bases constructed 
in~\cite{Beneke:2009rj}, one can check that this identity is in fact verified.
Coming finally to the real part of the NLL matrix elements, one finds from quark-quark 
scattering and gluon-gluon scattering respectively that the corresponding impact 
factors must be given by
\beq
 C_a^{(1)} \, = \, \frac{1}{2} Z_{1, {\bf R}, a}^{(1)}  + \frac{1}{2} \widehat{H}^{(1),0,[8]}_{aa} \, ,
\label{imp1}
\eeq
where we defined $\widehat{H}^{(1),0,[8]} \equiv H^{(1),0,[8]}/H^{0,[8]}$. Quark-gluon 
scattering is then completely determined, if universality has to hold. This is easily verified 
to be true for divergent terms, using the explicit expressions for ${\cal Z}_{\bf 1, R}^{(i)}$. 
For finite contributions one needs
\beq
 {\rm Re} \left( \widehat{H}^{(1),0,[8]}_{qg} \right) \, = \, \frac{1}{2} \left[ {\rm Re} \left( 
 \widehat{H}^{(1),0,[8]}_{gg} \right) + {\rm Re} \left( \widehat{H}^{(1),0,[8]}_{qq} \right) 
 \right] \, ,
\label{ReNLL1}
\eeq
which can also be verified using the one-loop results derived in~\cite{Kunszt:1993sd}.

\section{Results at two loops and beyond}
\label{resu}

The results presented in \secn{pertexp} are of course not new, and serve mostly to illustrate
the method we use. Things get more interesting starting at two loops. At this level, the LL
results simply confirm the exponentiation implicit in the factorized expression~(\ref{ReggeFact}).
On the other hand the real part of the single-logarithmic NLL matrix elements provide an 
expression for the two-loop Regge trajectory,
\beq
 \alpha^{(2)}  \,  =  \, C_A K^{(2)} + \RE \left[ \widehat{H}^{(2), 1, [8]}_{ab} \right]  + \ord(\eps) \, ,
\label{alpha2}
\eeq
which again corresponds both to expectations and to known results. Yet more interesting
are the results for impact factors obtained by fitting the real parts of the NNLL matrix elements.
Indeed one finds
\beqa
  C_a^{(2)} & = & \frac{1}{2} Z^{(2)}_{1, {\bf R}, aa} - \frac{1}{8} 
  \left( Z^{(1)}_{1, {\bf R}, aa} \right)^2 + \frac{1}{4} Z^{(1)}_{1, {\bf R}, aa} \, 
  \RE \left[ \widehat{H}^{(1), 0, [8]}_{aa} \right] - \frac{1}{4} R_{aa}^{(2), 0, [8]}
  \label{c2imp} \\ && \hspace{-17mm} 
  - \frac{\pi^2 (K^{(1)})^2}{4} \bigg\{ \left[ \left( {\bf T}^2_{s, aa} \right)^2 \right]_{[8], [8]}
  - {\cal C}_{{\rm tot}, aa} \left[ {\bf T}^2_{s, aa} \right]_{[8], [8]}
  + \frac{1}{4} {\cal C}^2_{{\rm tot}, aa} - \frac{(1 + \kappa_{aa}) C_A^2}{2} \bigg\} 
  \, + {\cal O} \left( \epsilon^0 \right) \, . \nonumber 
\eeqa
If high-energy factorization were to be exact at this order, the remainders $R_{aa}^{(2), 0, [8]}$
would vanish. One sees immediately however that this cannot work: indeed, while the first
line of \eq{c2imp} would have the proper degree of universality, the second line involves
both mixing of color components through the $s$-channel color operator ${\bf T}_s$, as 
well as process-dependent terms. We conclude that Regge factorization as embodied in 
\eq{ReggeFact} must break down, and \eq{c2imp} provides us with the tools to isolate
the factorization-breaking terms. We can define universal impact factors using the first
line of \eq{c2imp}, as
\beq
  \widetilde{C}_a^{(2)} \, = \, \frac{1}{2} Z^{(2)}_{1, {\bf R}, aa} - \frac{1}{8} 
  \left( Z^{(1)}_{1, {\bf R}, aa} \right)^2 + \frac{1}{4} Z^{(1)}_{1, {\bf R}, aa} \, 
  \RE \left[ \widehat{H}^{(1), 0, [8]}_{aa} \right] \, + {\cal O} \left( \epsilon^0 \right) \, ,
\label{newC}
\eeq
while the second line naturally defines non-factorizing remainders $\widetilde{R}^{(2), 
0, [8]}_{ab}$. We can check the consistency of our approach at the two-loop level: first 
we can compute the newly defined remainders for quark and gluon amplitudes by using 
the explicit color bases of Ref.~\cite{Beneke:2009rj}, with the results
\beq
  \widetilde{R}^{(2), 0, [8]}_{qq} \, = \, \frac{\pi^2}{4 \epsilon^2}
  \left(1 - \frac{3}{N_c^2} \right) \, , \quad \, 
  \widetilde{R}^{(2), 0, [8]}_{gg} \, = \,  - \, \frac{ 3 \pi^2}{2 \epsilon^2} \, , \quad \,
  \widetilde{R}^{(2), 0, [8]}_{qg} \, = \, - \, \frac{\pi^2}{4 \epsilon^2}
  \,. \nonumber
\eeq
Next, we can construct a function measuring the discrepancy between the predictions 
of Regge factorization for the quark-gluon amplitude, which are based on universality, 
and the actual matrix elements. We find
\beqa
  \Delta_{(2),0,[8]} & \equiv & \frac{M^{(2),0}_{qg}}{H^{(0),[8]}_{qg}} - 
  \bigg[C^{(2)}_q + C^{(2)}_g + C^{(1)}_q C^{(1)}_g - \frac{\pi^2}{4} 
  \left(1 + \kappa \right) (\alpha^{(1)})^2 \bigg] \nonumber \\
  & = & \frac{1}{2}\bigg[\widetilde{R}^{(2), 0, [8]}_{qg} - \frac{1}{2} 
  \left(\widetilde{R}^{(2), 0, [8]}_{qq} + 
  \widetilde{R}^{(2), 0, [8]}_{gg} \right)\bigg] 
  \, = \, \frac{\pi^2}{\eps^2} \frac{3}{16} \left(\frac{N_c^2 + 1}{N_c^2} \right)\, ,
\label{delta}
\eeqa
which precisely reproduces the result of Ref.~\cite{DelDuca:2001gu}.

Clearly, soft-collinear factorization can be used to identify precisely, order by order 
in perturbation theory, the non-universal terms that break Regge factorization. Moving
on to the three-loop level, one finds, as expected, that the breaking of universality
occurs at the level of single-logarithmic terms. Indeed, if one attempts to find an 
expression for the three-loop Regge trajectory using soft-collinear ingredients one
finds
\beqa
  \alpha^{(3)} & = & C_A K^{(3)} + \frac{\pi^2 (K^{(1)})^3}{2} 
  \bigg[{\cal C}_{{\rm tot}, ab} N_c \left( {\bf T}^2_{s, ab} \right)_{[8],[8]}
  - \frac{{\cal C}^2_{{\rm tot}, ab} N_c}{4} + \frac{1 + \kappa_{ab}}{2} N_c^3 \nonumber  \\
  && \, - \, \frac{1}{3} \sum_n \left(2 N_c + {\cal C}_{[n]} \right) \left| 
  \left({\bf T}^2_{s, ab} \right)_{[8], n} \right|^2 \bigg] -\frac{1}{2} R^{(3), 1, [8]}_{ab}
  + {\cal O} \left( \epsilon^{-2} \right) \, ,
\label{alpha3}
\eeqa
where lower-order poles can also be determined~\cite{us} but are omitted here for brevity.
Once again, the first term in \eq{alpha3} is universal and has the expected form, while
the second term involves both color mixing and process-dependent contributions. Following
our general strategy, we define
\beq
 \widetilde{\alpha}^{(3)} \, = \, K^{(3)} N_c + {\cal O} \left( \epsilon^0 \right) \, .
\label{traj3}
\eeq
We can now determine the three-loop, single logarithmic remainders, by computing the
non-universal contributions to \eq{alpha3} in the color bases of Ref.~\cite{Beneke:2009rj}.
The results for quark and gluon amplitudes are
\beq
  \widetilde{R}^{(3), 1, [8]}_{qq} \, = \, \frac{\pi^2}{\epsilon^3} \, 
  \frac{2 N_c^2 - 5}{12 N_c} \, , \quad
  \widetilde{R}^{(3), 1, [8]}_{gg} \, = \, - \, \frac{\pi^2}{\epsilon^3} \,
  \frac{2}{3} \, N_c \, , \quad
  \widetilde{R}^{(3), 1, [8]}_{qg} \, = \, - \, \frac{\pi^2}{\epsilon^3} \,
  \frac{N_c}{24} \, .
\label{explR3}
\eeq
Together with \eq{ReggeFact}, \eq{explR3} constitutes an explicit prediction for the
high-energy limit of quark and gluon amplitudes at three loops. 

We conclude by noting that, just as soft-collinear factorization provides important 
information in the high-energy limit, one may also use Regge factorization to extract
constraints on the finite parts of the amplitudes, which are not in principle controlled by 
infrared physics. An example of this was discussed at the one-loop level in \secn{pertexp},
most notably in Eqns.~(\ref{ImNLL1}) and (\ref{magid}): there, the reality of the Regge
trajectory and impact factors was reflected in properties of the one-loop hard imaginary 
parts. It turns out that it possible to generalize these constraints to all orders in 
perturbation theory, using extensions of the color identity given in~\eq{magid}.
For the LL and NLL hard-scattering coefficients we find~\cite{us}
\beqa
  {\rm Im} \left( \widehat{H}^{(n), n, [8]} \right) & = & 0 \, , \nonumber \\
  {\rm Re} \left( \widehat{H}^{(n), n, [8]} \right) & = & \frac{1}{n!}
  \left( \widehat{H}^{(1), 1, [8]} \right)^n \, = \, O(\epsilon^n) \, , \label{allord} \\
  {\rm Im} \left( \widehat{H}^{(n), n - 1, [8]} \right) & = & - \pi \, \frac{1 + \kappa}{2}
  \left( n \, \widehat{H}^{(n), n, [8]}\right)  \, = \, O(\epsilon^n) \, , \nonumber \\
  {\rm Re} \left(\widehat{H}^{(n), n - 1, [8]} \right) & = & {\rm Re} \left( 
  \widehat{H}^{(2), 1} \right) \widehat{H}^{(n - 2), n - 2} + (2 - n) \, {\rm Re} 
  \left( \widehat{H}^{(1), 0, [8]} \right) \widehat{H}^{(n - 1), n - 1} 
  \, = \, {\cal O} (\epsilon ^{n-2}) \, . \nonumber 
\eeqa
In essence, \eq{allord} reinforces the idea that high-energy logarithms are in fact 
infrared in nature: indeed, leading and next-to-leading logarithmic contributions
to hard scattering coefficients are forced to vanish with increasing powers of
the regulator $\epsilon$. This means that infrared-finite high-energy logarithms
must come from the interference of soft and collinear functions with lower-order
contributions subleading in $\epsilon$. These constraints, discussed in greater 
detail in~\cite{us}, are the subject of ongoing investigations.

\vspace{1cm}

{\large{\bf Acknowledgements}}

\vspace{2mm}

\noindent This work was supported by MIUR (Italy), under contract 2010YJ2NYW$\_$006; 
by the University of Torino and by the Compagnia di San Paolo under contract ORTO11TPXK; 
by the Research Executive Agency (REA) of the European Union, through the Initial Training 
Network LHCPhenoNet under contract PITN-GA-2010-264564, and by the ERC grant 291377 
``LHCtheory: Theoretical predictions and analyses of LHC physics: advancing the precision 
frontier''.

\end{document}